\documentclass[12pt,preprint]{aastex}

\newcommand{\eg}{{\it e.g.}}

\newcommand{\etal}{{\it et~al.}}

\newcommand{\evts}{\ensuremath {\:{\rm events\:cm^{-2}\:s^{-1}}}}
\newcommand{\cps}{\ensuremath {\:{\rm counts\:pixel^{-1}\:s^{-1}}}}

\newcommand{\pden}{\ensuremath{\:{\rm particles\:cm^{-2}\:s^{-1}}}}
\newcommand{\cerenkov}{\v{C}erenkov}
\newcommand{\kalpha}{\ensuremath {{\rm K_{\alpha}}}}
\newcommand{\kbeta}{\ensuremath {{\rm K_{\beta}}}}

\newcommand{\wfirst}{{\it WFIRST\/}}
\newcommand{\sdo}{{\it SDO\/}}
\newcommand{\re}{\ensuremath{R_E}}
 
\newcommand{\myemail}{Jeffrey.W.Kruk@nasa.gov}

\slugcomment{Accepted by PASP on November 10, 2015}

\shorttitle{Instrumental Backgrounds in Geo-synchronous Orbits}
\shortauthors{Kruk et al.}

\begin{document}

\title{Radiation-Induced Backgrounds in Astronomical Instruments: Considerations for Geo-synchronous Orbit and Implications for the Design of the WFIRST Wide-Field Instrument
}

\author{Jeffrey W. Kruk, Michael A. Xapsos, Nerses Armani, Craig Stauffer}
\affil{NASA Goddard Space Flight Center, Greenbelt, MD 20771}

\and

\author{Christopher M. Hirata}
\affil{Center for Cosmology and AstroParticle Physics (CCAPP),
The Ohio State University, \\
191 West Woodruff Ave., Columbus, Ohio 43210}
    
\email{\myemail}

\begin{abstract}
Geo-Synchronous orbits are appealing for Solar or astrophysical observatories because they permit continuous data downlink at high rates. The radiation environment in these orbits presents unique challenges, however. This paper describes both the characteristics of the radiation environment in Geo-Synchronous orbit and the mechanisms by which this radiation generates backgrounds in photon detectors. Shielding considerations are described, and a preliminary shielding design for the proposed Wide-Field InfraRed Survey Telescope observatory is presented as a reference for future space telescope concept studies that consider a Geo-Synchronous orbit.
\end{abstract}


\section{Introduction}

Geo-synchronous orbits (circular orbits about the Earth with an orbit period equal to one sidereal day) have several attributes that make them attractive to Solar and astronomical observatories:  a satellite in such an orbit will remain near a fixed longitude above the Earth, enabling continuous contact with a ground station, and high data downlink rates can be provided by practical on-board transmitters.
The Solar Dynamics Observatory (\sdo) is an example of such a mission: it was launched into a 
28\degr\ Geo-synchronous orbit that provides nearly uninterrupted visibility of the Sun and continuous contact with a ground station in White Sands, New Mexico.  The combination of 24-hour per day continuous downlink and a total data rate of 150 Mbps were critical to achieving the scientific goals of the \sdo\ mission.
New RF systems presently under development are expected to enable even greater data volumes to be delivered from Geo-synchronous orbits. For this reason, the Wide-Field InfraRed Survey Telescope (\wfirst) mission has been investigating an orbit similar to that of \sdo\ \citep{sdt15}. The radiation environment in Geo-synchronous orbit, however, poses some interesting challenges to astronomical mission design.

The impact of the Geo-synchronous orbit radiation environment on electronics and mechanisms is significant, but long experience with communications satellites and scientific missions such as \sdo\ has resulted in design practices and rigorous testing regimes that are well-understood (see, \eg, \citeauthor{hdbk4002}). Discussion of these aspects of observatory design is outside the scope of this paper. Similarly, there is no discussion of the effects of radiation damage to instrumentation caused by long-term exposure to the space environment, as such effects are often unique to the details of the design of any given detector. Instead, the focus here is on understanding the impacts of the radiation environment on instrumental backgrounds, and how to minimize degradation of data quality. 

Some components of the Geo-synchronous orbit radiation environment are common to other high-Earth orbits or deep space. These include Galactic cosmic rays, particles in the Solar wind, and coronal mass ejections and Solar flares (collectively designated here as Solar particle events).  Galactic cosmic rays occur at relatively low flux levels, typically 1-4 \pden, but have high energies: the distribution peaks at 0.5-1\,GeV/nucleon  \citep[depending on phase of the Solar cycle - see Fig. 5 of][]{bx08} and extends up to a TeV. 

Because the photon detectors in common use in astronomical instruments work by collection of electron-hole pairs in semiconductors, or by detection of photo-electrons produced by photon interactions in a photo-cathode, they are also excellent detectors of energetic charged particles. Spurious signals generated in detectors by energetic charged particles in the space radiation environment can therefore degrade scientific data even if the detectors themselves are immune to long-term radiation damage. The energies of Galactic cosmic rays, and of the particles in the high-energy tail of the distribution of Solar particle events, are too high for shielding to be practical, so these particles set an irreducible floor on the rate of background events in astronomical detectors. These rates are generally low enough that it is practical to mitigate these backgrounds by a combination of observation procedures, limiting exposure durations and obtaining redundant exposures, and data processing algorithms that detect and mask out affected pixels when combining exposures.

The distinguishing characteristic of Geo-synchronous orbits is that they lie within the outer Van Allen belt. The outer belt is populated primarily with electrons, with densities of roughly $10^{7}$ \pden\ and energies extending to 8\,MeV and higher \citep{ae9ap9}.  This high-energy electron environment in Geo-synchronous orbit poses a difficult challenge, as the low mass of electrons leads to shielding considerations that are qualitatively different from those of protons or heavy ions, and the high particle density means that the standard mitigation strategies mentioned above are not adequate in themselves for most applications.

This paper will explore the ways in which the Geo-synchronous orbit environment can affect common astronomical instruments, and present the preliminary shielding design concept developed for the \wfirst\ wide-field instrument as a concrete example of how these effects can be avoided or mitigated. Two detector materials will receive particular attention in sample calculations: silicon, because of its wide use, and HgCdTe, because it will be the material used in the \wfirst\ wide-field instrument.

This paper has two major goals: (i) to assess the impact of high-energy particles on WFIRST observations in GEO; and (ii) to provide a record of the design process and issues encountered as a reference for future space observatory concept studies that consider a Geo-synchronous orbit. The intended audience includes both astronomers who may be familiar with data processing techniques for cosmic ray detection but not with charged particle propagation, and engineers who may be familiar with designing spacecraft subsystems for surviving space radiation environments but not familiar with how radiation backgrounds can degrade data quality. Thus the paper includes some material that is tutorial in nature in order to provide a discussion that is reasonably self-contained.

The organization of this paper is as follows. Section 2 provides general information on relevant aspects of the interaction of radiation with common structural, optical, and detector materials; Section 3 provides representative estimates of the radiation environment within an instrument; Section 4 provides a concrete example of a shielding concept developed for the proposed \wfirst\  mission; and Section 5 presents concluding remarks.

\section{Physical principles for shielding and particle interactions in detectors.}

\subsection{Charged-particle interactions in detector materials}
Photon detectors in common use in astronomy often take the form of photodiode arrays. These are solid state devices in which a reverse bias is applied to photo-sensitive diodes. When an optical or NIR photon is absorbed in the diode, an electron-hole pair is produced. The reverse bias separates the electron and hole, preventing recombination, and causes the charges to be stored in the diode. 
The accumulated signal is read out at the end of an exposure and is proportional to the incident photon flux.

The detector technology is similar for X-ray and gamma ray detectors, though the higher energy of the photons results in some operational differences: the active layer is often much thicker, and the signal level is much higher: instead of a single electron-hole pair, the energetic photo-electron produces thousands of electron-hole pairs via ionization. This signal is much greater than the electronic noise, and the incident fluxes are generally low enough that continuous readouts provide photon counting with the signal being proportional to the energy of each incoming photon. Thus the interaction of a charged particle will be registered as a single photon detection.
 Examples include the thick deep-depletion CCDs used for X-ray detection in the EPIC instrument on {\it XMM} or the {\it Swift} XRT \citep{Holl96}, and the CdZnTe detector pixels in the {\it Swift} Burst Alert Telescope for gamma-ray detection \citep{Bart05}.

Some detectors in common use have very different architectures that do not involve photodiode arrays, such as micro channel plate (MCP) image intensifiers used at ultraviolet wavelengths (\eg\ the FUV detector in HST/COS and the MAMA detectors in HST/COS and HST/ACS), but these too are sensitive to charged particles and would be affected by the charged particle backgrounds discussed in this paper. These detectors are also ordinarily operated in photon-counting mode.

\begin{figure}[htb!]
\epsscale{1.0}
\plotone{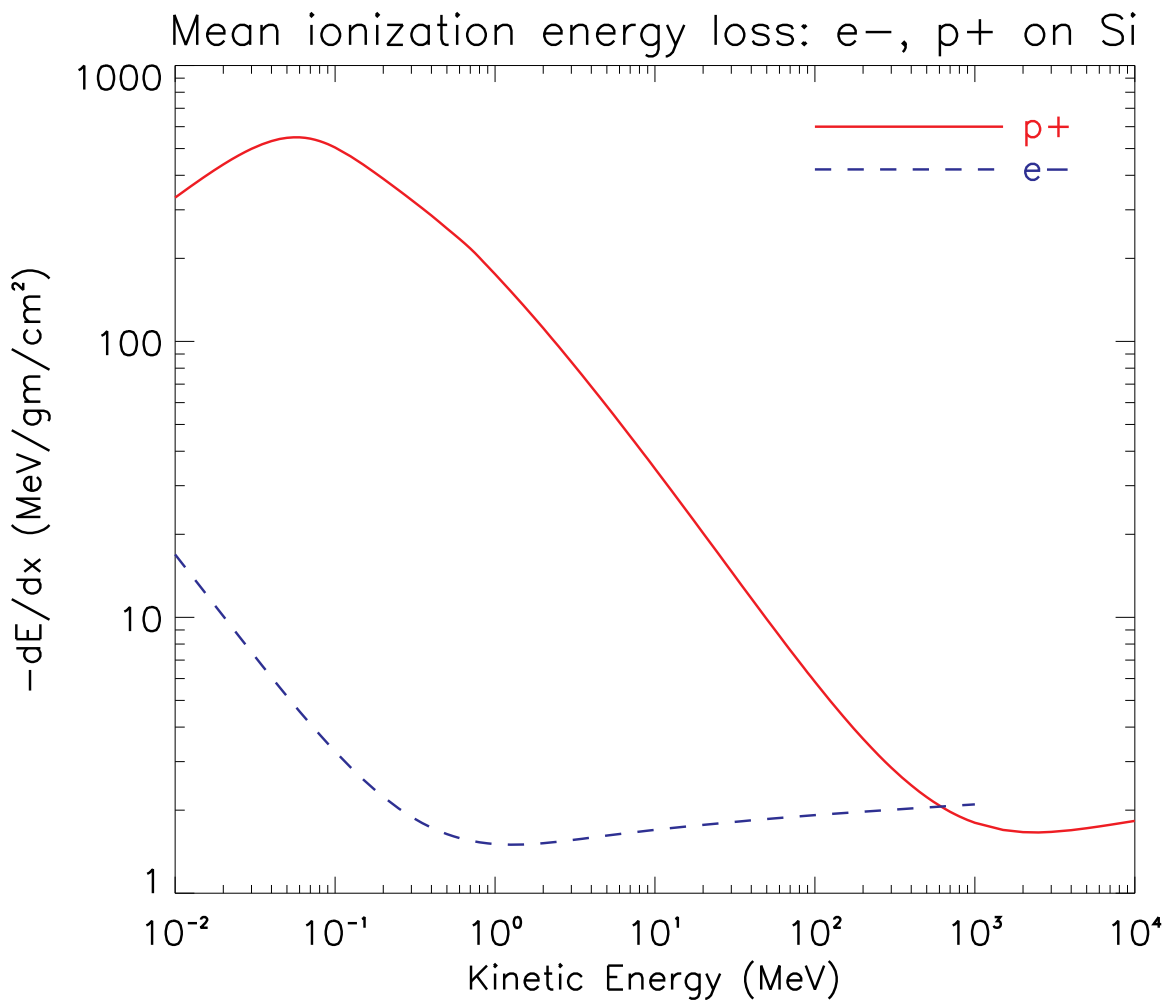}
\caption{The mean energy loss due to ionization for electrons (dashed line) and protons (solid line) passing through silicon is plotted as a function of kinetic energy. The curve for electrons is similar to that for protons, except for being displaced a factor of $\approx$ 2000 to lower energies. Electrons thus lose much less energy per unit path length than protons at energies of interest here.}\label{fig-dedx_si}
\end{figure}

The primary process by which an energetic heavy particle interacts with matter is by ionization of the atoms in the vicinity of its path. This process is described in many text books; a useful review is included in the biannual Review of Particle Properties (\eg\ \citet{ber12}, and online at \url{http://pdg/lbl.gov}).
The mean ionization energy loss per unit density-weighted path length is given by the Bethe-Bloch equation; discussion of this formula and tabulations of parameters for a wide range of materials is provided by \citet{sb82}, \citet{sb84}, and \citet{sbs84}.
Evaluation of the ionization energy loss for electrons and protons in silicon is shown in Figure \ref{fig-dedx_si} as an example. The shape of the curve is similar for all materials, and the energy-dependence is a complex function of the velocity of the particle. This velocity-dependence causes the curve for electrons to be displaced by a factor of $\approx$2000 to lower energies relative to that for protons. Particles with energy well above that of the minimum in the curve lose energy relatively slowly and can traverse large path lengths; however, once the energy falls below the minimum in the curve, losses relative to the energy of the particle grow rapidly and little additional material is needed to stop the particle. The majority of protons encountered in Geo-synchronous orbit have energies corresponding to the steep part of the curve, while a substantial fraction of the electron population have energies for which ionization losses are small.

The minimum in the mean ionization energy loss curve for protons on silicon is 1.66\,MeV/gm/cm$^{2}$. The creation of an electron-hole pair in Si requires 3.63 eV, and a typical backside-illuminated silicon CCD detector might have a thickness of 15 microns. Thus: a minimum ionizing proton at normal incidence would deposit 5800 eV, resulting in a signal of 1600 electrons. The deposited energy grows only slowly for higher proton energies, so this is a representative value for the signal resulting from passage of Galactic cosmic rays. For lower-energy protons, such as the trapped protons encountered in the SAA, the signal deposited in pixels can be significantly larger.

Detectors that integrate the incoming photons over long periods of time, as is common for solid-state detectors at UV, visible, and near-IR wavelengths, are ordinarily operated so that a detected photon generates a single electron-hole pair. In such detectors, the large charge deposition by charged particles is equivalent to receipt of hundreds or thousands of photons, which often leads to significant corruption or effective total loss of the astronomical signal in a given pixel for the exposure in progress. For photon-counting detectors, however, a charged particle interaction is typically registered as a single spurious photon detection. For many applications, the number of such background events that can be tolerated may be much greater than for integrating detectors.

\subsection{Shielding of charged particles}

Instrument detector systems are rarely exposed directly to the orbital radiation environment, but rather are commonly deeply embedded in the interior of a satellite. Structural materials, optical elements, electronics boxes, etc, often provide a mass-density sufficient to block all but high energy protons over large portions of the total solid angle. The remainder of the solid angle, however, is often covered by little more than multi-layer insulation and thin panels designed for light baffling, thermal control, and contamination control. Once the layout of spacecraft and payload components has been established, designers can assess if it would be beneficial to include additional material to increase shielding in certain directions.

Because most of the ionization energy loss results from scattering by atomic electrons, and incident particles such as protons or heavy atomic nuclei have much higher masses than electrons, their trajectories deviate only slightly as a result of these collisions. Thus, when designing shielding for protons and heavy ions, only the total line-of-sight mass density matters (with minor variations for different materials).

\begin{figure}[htb!]
\epsscale{1.0}
\plotone{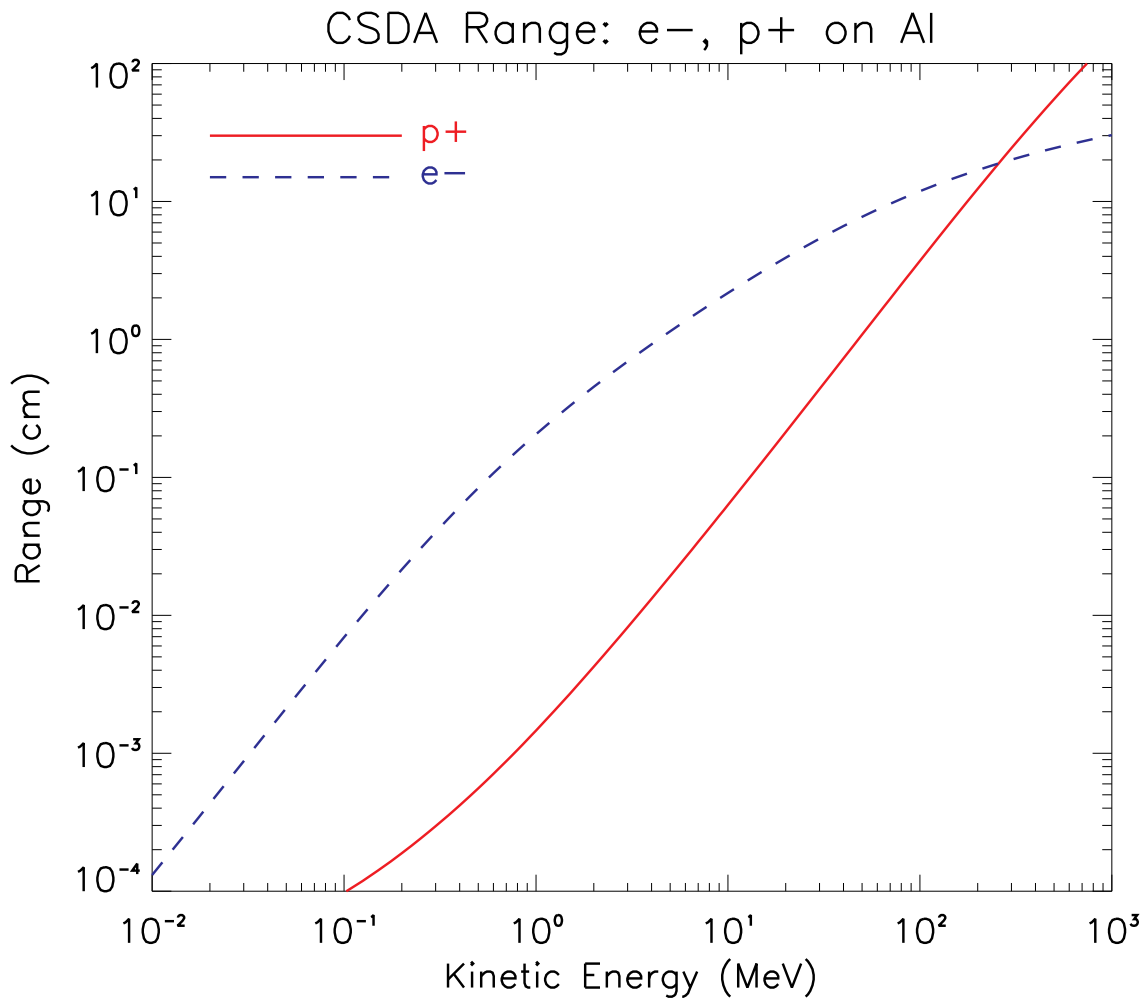}
\caption{The CSDA range of electrons (dashed) and protons (solid) passing through Aluminum is plotted as a function of kinetic energy. The greater range of electrons at low energies is apparent. At high electron energies, energy loss to bremsstrahlung dominates over ionization and the slope of the range curve decreases.}.\label{fig-prange}
\end{figure}

The range of a particle in a material is ordinarily calculated in the Continuous Slowing Down Approximation (CSDA), in which the inverse of the mean energy loss is integrated over energy from zero to the incident energy of the particle. This has been calculated for both electrons and protons incident on aluminum to illustrate the shielding provided by typical instrument enclosures, and is shown in Figure \ref{fig-prange}. A convenient reference point that illustrates the scale for shielding considerations is that a 50\,MeV proton has a range just slightly greater than one centimeter of aluminum. Similarly, the typical 2.5mm aluminum panel corresponds roughly to the range of a 22\,MeV proton at normal incidence. As can be seen from the plot, as proton energies increase beyond 50\,MeV, the shielding thickness required to stop them grows rapidly. The mass required for such shielding is usually prohibitive for a space mission, unless the area to be covered is quite small. Conversely: once the energy of the particle has dropped below the minimum of the energy loss curve, the rapid rise in energy loss with decreasing energy causes the range to drop very quickly.
The range for electrons is much greater than that for protons at energies under $\approx$200\,MeV, but the electron energies of interest here are under 10\,MeV and thus have a range of a few cm at most. This will be explored in more detail in the following sections.

\section{The Orbital Radiation Environment }

Reviews of the near-Earth radiation environment can be found in \citet{fg96}, \citet{bds03}, and \citet{bx08}, and a recent review of updates to models of this environment is given by \citet{xoo13}. A brief summary of the proton and electron environment in geo-synchronous orbit is presented here, but the reviews listed above and references therein should be consulted for a more extensive discussion. In particular, heavy ions are not discussed further because they are relatively few in number and thus are not significant contributors to backgrounds in the photon detectors of interest here.

\subsection{Protons}
\begin{figure}[htb!]
\epsscale{1.0}
\plotone{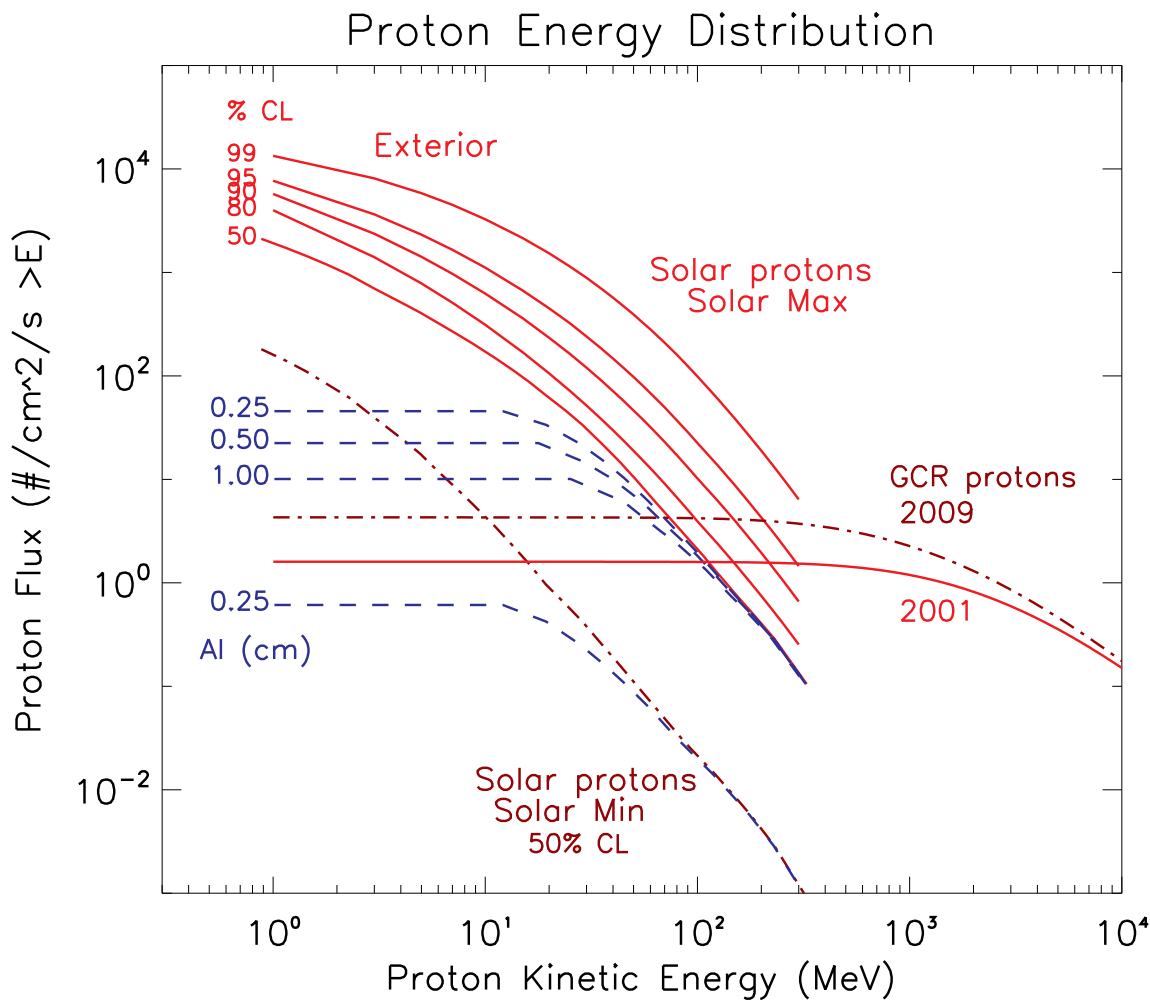}
\caption{The integral energy distributions of protons in Geo-synchronous orbit are plotted for a range of conditions. The monthly-averaged energy distribution of Galactic cosmic ray protons is shown for January 2001 and January 2009, as estimated by the Badhwar-O'Neill 2014 model. These time periods are representative of the minimum and maximum fluxes at the orbit of the Earth over the past Solar cycle. Annual-average Solar particle event protons are plotted for a variety of Solar Maximum ({\it solid lines}) and typical Minimum ({\it dot-dash line}) conditions. The Confidence Levels (CL) give the probability that the actual flux distribution received in any given year will be less than that shown.  Also shown is the distribution for Solar protons in typical years after passing through various thicknesses of aluminum ({\it dashed lines}). For Solar maximum conditions, a significant thickness of material is required to reduce their number to below that of the Galactic cosmic rays. } \label{fig-swp}
\end{figure}

The proton environment in Geo-synchronous orbit consists of the Solar particle events and Galactic cosmic rays (GCR). 
The GCR flux is modulated by the Solar magnetic field over the course of the Solar cycle; representative integral energy distributions from the BON2014 GCR model \citep{bon15} are shown in Figure \ref{fig-swp}. 
The GCR proton fluxes range from 1.6 to 4.3 \pden. 
The differential flux distribution peaks at about 300\,MeV and 700\,MeV, for  2009 and 2001, respectively; these energies are too high for shielding to be practical. Attempting to do so is likely to be counterproductive, as the probability of hadronic showers increases and the net effect is an increase in the number of pixels hit.

Trapped protons and the majority of Solar wind protons are at low enough energies that they will not penetrate typical spacecraft or instrument enclosures.  However, protons arising from Solar particle events can have high energies.  The number and intensity of such events varies significantly from year to year.  The variation in the annual-average energy distribution of Solar particle event protons, taken from the Emission of Solar Protons model \citep[ESP,][]{esp00}, is shown in Figure \ref{fig-swp}.   
The Solar maximum fluxes are shown for Confidence Levels (CL) ranging from 50\% to 99\%, which indicates the probability that any given year will have a lower flux than that shown.
For the median (CL=50\%) Solar Maximum energy distribution, the number of protons that can penetrate even 1cm of aluminum is comparable to the GCR rate, and the CL=95\% distribution is roughly an order of magnitude greater than the CL=50\% distribution. 

\begin{figure}[htb!]
\epsscale{1.0}
\plotone{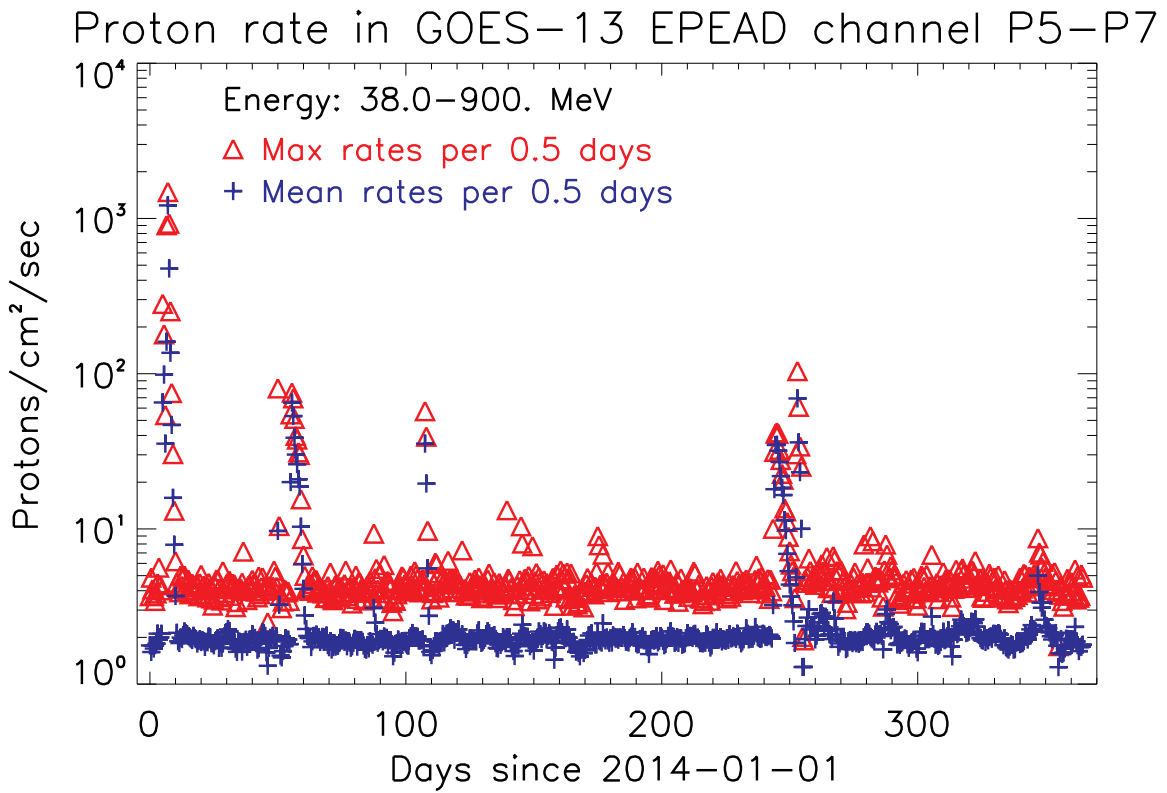}
\caption{The proton flux measured by the GOES-13 EPEAD instrument is plotted  vs. time for 2014, for $E_{proton} \geq$ 38\,MeV. 
For most days, the mean and maximum rates are 2 and 4  \pden, respectively, but for a small number of days the rates are orders of magnitude higher. 
}\label{fig-prp}
\end{figure}

It is important to note that the high-energy Solar proton flux is highly variable on short time scales, not just on an annual basis: most of the flux will be experienced in relatively short periods of time. This variability is illustrated in  
Figure \ref{fig-prp}, which shows the energetic proton flux density in Geo-synchronous orbit as a function of time in the year 2014. The points on the plot represent the mean and maximum 5-minute average fluxes for half-day periods throughout the year. Even on relatively quiet days, the maximum 5-minute average flux is twice the mean, and for a handful of days the fluxes are orders of magnitude higher than for a typical day. However, if the nature of the observing program is such that short periods of high backgrounds can be tolerated, then one doesn't need to design shielding for the times of peak fluxes. 
 
The variability of the Solar proton environment on short time-scales has important implications for an instrument designer. If the mission objectives require continuous availability of high-quality data, then shielding design and choice of detector architecture may have to be based not just on a proton spectrum corresponding to one of the high confidence level curves in Figure \ref{fig-swp}, but also on further scaling of this spectrum to account for the fact that it is delivered on short time scales. However, if the mission objectives allows degraded data quality for a modest fraction of the mission duration, then instruments may be designed based on a more benign proton distribution corresponding to the required fraction of good-quality data.

\subsection{Electrons}

The outer Van Allen belt is a toroidal region extending roughly from 3 to 10 Earth Radii (\re),with greatest particle intensities at 4-5 \re. The radius of a Geo-synchronous orbit corresponds to 6.6 \re, just outside the regions of highest particle intensity. The outer belt is populated primarily with electrons, with densities of roughly $10^{7}$ \pden\ and energies extending to 8\,MeV and higher \citep{ae9ap9}.  
These electron populations are highly variable and can intensify by orders of magnitude over a time period of a few days.  Electrons are generally the dominant charge carriers at most shielding depths in geosynchronous orbits.  They are important to consider for evaluating risk for total ionizing dose, displacement damage and charging effects. Protons in the outer belt are numerous, but their maximum energy is significantly lower.

\begin{figure}[htb!]
\epsscale{1.0}
\plotone{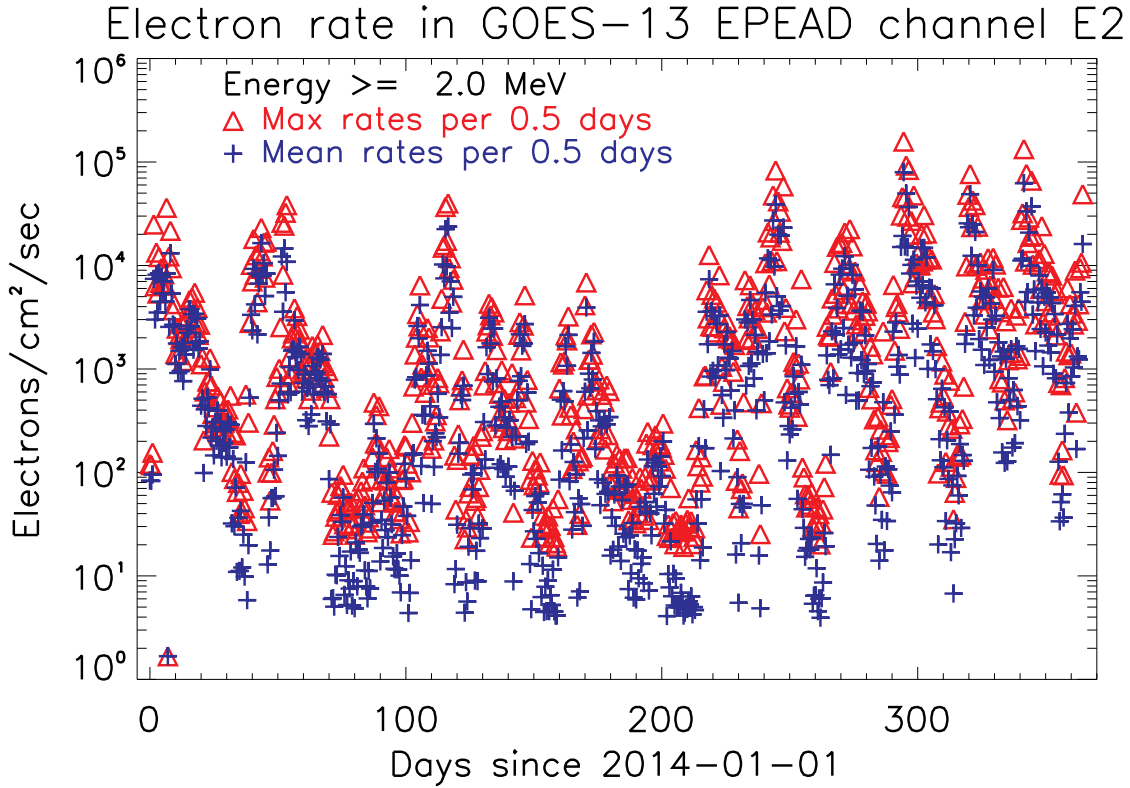}
\caption{The electron flux measured by the GOES-13 EPEAD instrument is plotted as a function of time for 2014, for $E_{electron} \geq $ 2\,MeV.
There is substantial time variability, with rates varying by 3 orders of magnitude over intervals of a few days to a month. }\label{fig-erp}
\end{figure}

The time-variability of the Geo-synchronous orbit trapped-electron environment is illustrated in Figure \ref{fig-erp}, and an annual-average energy spectrum is shown as the solid black line in the upper right panel of Figure \ref{fig-AlBox}. The electron flux is far more variable than for protons, and there are very few time periods when the electron flux might be deemed to be ``quiet.'' The electrons are also far more numerous than the protons at energies below a few MeV, but the energy distribution drops more steeply so that there are few electrons at energies greater than 10\,MeV.

\subsubsection{Propagation}
The low mass of electrons causes the propagation of electrons in materials to differ from that of protons in a number of important respects. The underlying physics of the interactions is the same, but effects that are tiny corrections for protons at the energies of interest become  large in the case of electrons. 

\begin{figure}[htb!]
\epsscale{1.0}
\plottwo{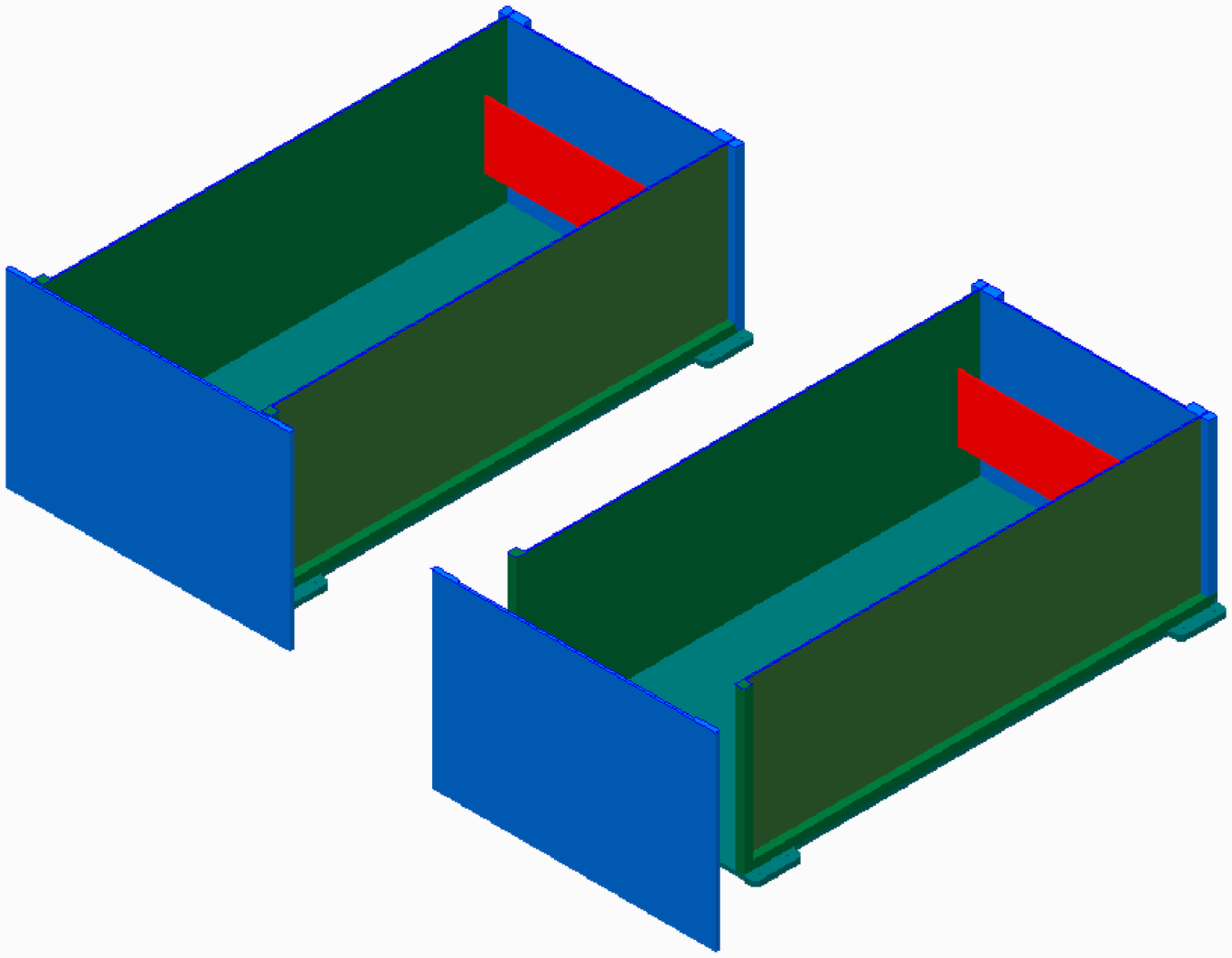}{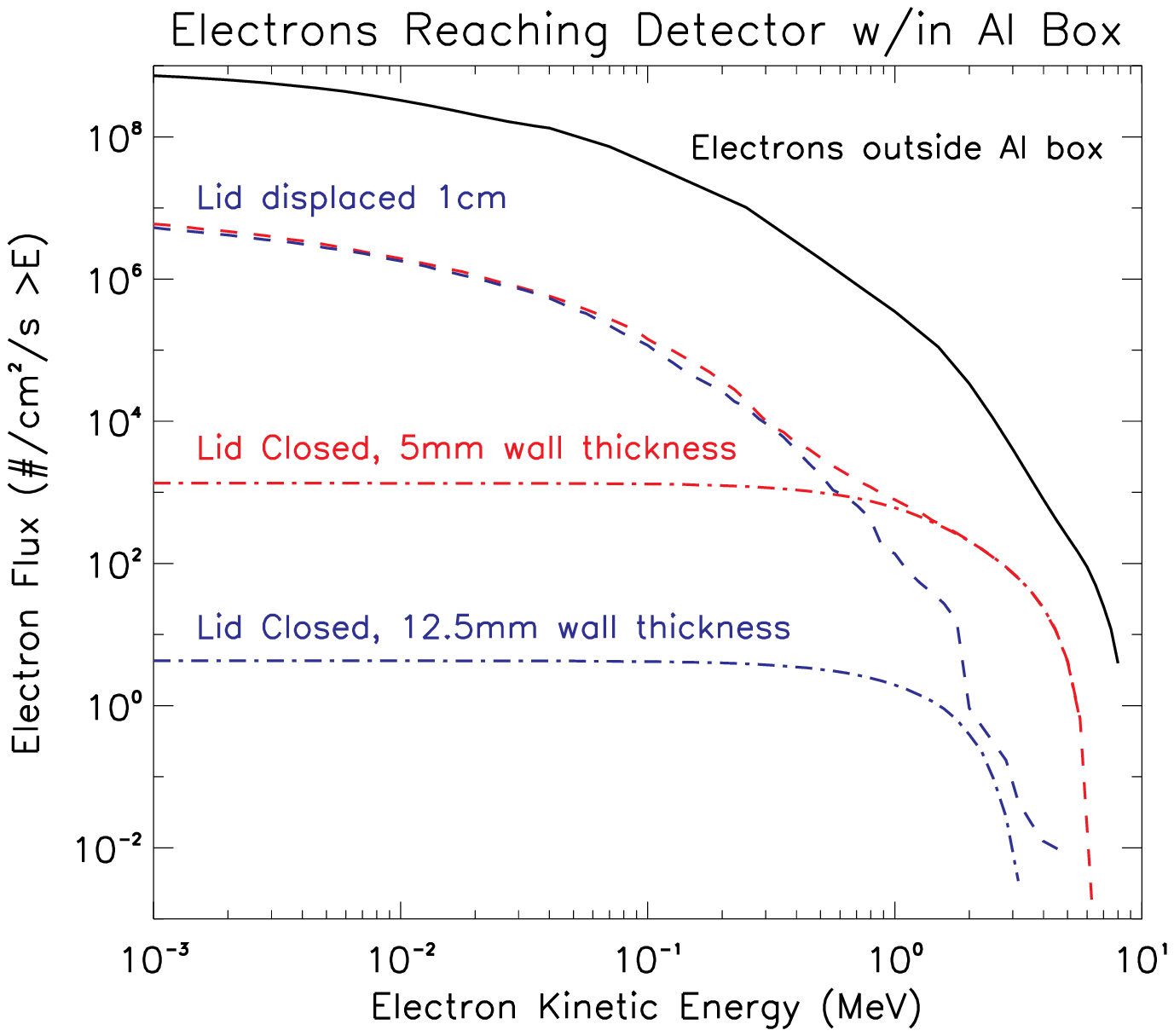}
\caption{This figure illustrates the propagation of electrons into aluminum boxes of two wall thicknesses and two geometries: one with the box sealed on all sides, and one with the lid displaced  along the axis of the box ({\it left panel}). The right panel gives the integral flux densities for the different configurations. The solid black line gives the annual average electron integral flux density in Geo-synchronous orbit (which is many orders of magnitude higher than the proton flux, even at Solar Max). The other lines show the integral fluxes reaching a detector surface at the end of the box opposite from the lid. The dot-dashed lines give the fluxes for the case of the sealed lid and the dashed lines give the fluxes for the displaced lid. The lid is oversized, so that there is no direct path from the outside to the detector. For energies below $\approx$ 2\,MeV for the thick wall and 0.7\,MeV for the thin wall, the fluxes are dominated by electrons that pass through the gap between the lid and the sides, and subsequently bounce around the interior of the box before reaching the detector.}\label{fig-AlBox}
\end{figure}

The first of these effects is the difference in scattering kinematics as electrons propagate through a material. The primary mechanism for energy loss by ionization is scattering from atomic electrons in the material, but deviations in the particles trajectories is dominated by scattering from nuclei \citep{Mol47, Lynch91}. The scattering is strongly forward-peaked for protons and heavy ions, with  angular deviations of trajectories on the order of a few degrees for energies of interest here. As a consequence, detector shielding design can be simplified by considering only the total line-of-sight material thickness between the detectors and the exterior of the satellite. Electrons, however, are $\approx$2000 times lighter than protons, so they will scatter at much larger angles. 
The importance of electron diffusion is illustrated by the simulation of electron propagation into a hollow aluminum box shown in Figure \ref{fig-AlBox}.
A detector was located at one end of the box, and the wall of the box opposite of the detector was either placed directly against the other walls to form a sealed enclosure, or displaced  outwards so that there was a gap around all 4 sides. The transverse dimensions of the lid were such that at least two bounces would be needed for a particle passing through the gap to reach the detector. The incident electron spectrum was a 1-year average for the Geo-synchronous orbit environment from the AE9 model \citet{ae9ap9}. The box geometry is illustrated in the left panel of 
Figure \ref{fig-AlBox}, and   
the results of a Monte-Carlo simulation using the NOVICE code (see below) are shown in 
the right panel,  
for wall thicknesses of 5\,mm and 12.5\,mm.  There is a dramatic difference in fluxes reaching the detector for the lid-displaced and lid-closed configurations. In the latter, there is substantial attenuation of the incident flux and only the highest-energy particles penetrate the walls. In the former, the attenuation is only a factor of $\approx$100: the distribution is dominated by particles passing through the opening in the side walls at the far end of the box and diffusing within the box until they reach the detector plane. Only for electrons of 1\,MeV and above, for the 5\,mm wall, is the distribution dominated by electrons penetrating through the walls themselves. For protons or other heavy particles, the lid-displaced distribution would have been essentially identical to the lid-closed configuration. The implication is that shielding for electrons must form a nearly-hermetically sealed enclosure around detectors.

Because optical systems require a clear light path from the celestial objects being studied to the detector, shielding electrons can be troublesome. Refractive materials can be used to allow light through while blocking electrons, but they must be thick enough to stop the electrons while still meeting all the optical requirements of the instrument. At far and extreme ultraviolet wavelengths, for example, there are no refractive materials that transmit the wavelengths of interest.

\subsubsection{Bremsstrahlung}
Another consequence of the low mass of electrons is the significant amount of X-ray and gamma-ray  photons \citep[bremsstrahlung or ``braking radiation'', see \eg][]{km59} produced as the electrons pass through materials.  While only a small fraction of the energy goes into bremsstrahlung photons (e.g. 1\% of the ionization energy loss for $\approx$0.7\,MeV electrons in Al, growing to 10\% of the ionization energy loss at 6\,MeV), these photons are much more penetrating than the primary electrons and can dominate backgrounds deep inside an instrument. The bremsstrahlung production cross-section scales as $Z^{2}$, where $Z$ is the atomic number of the material, so the composition of the observatory structure and the choice of shielding material can have a significant influence on the intensity of the X-ray background generated by the electrons. Quantitative estimates for the bremsstrahlung background in the case of the \wfirst\ wide-field instrument will be given in the next section.

\begin{figure}[htb]
\epsscale{1.0}
\plottwo{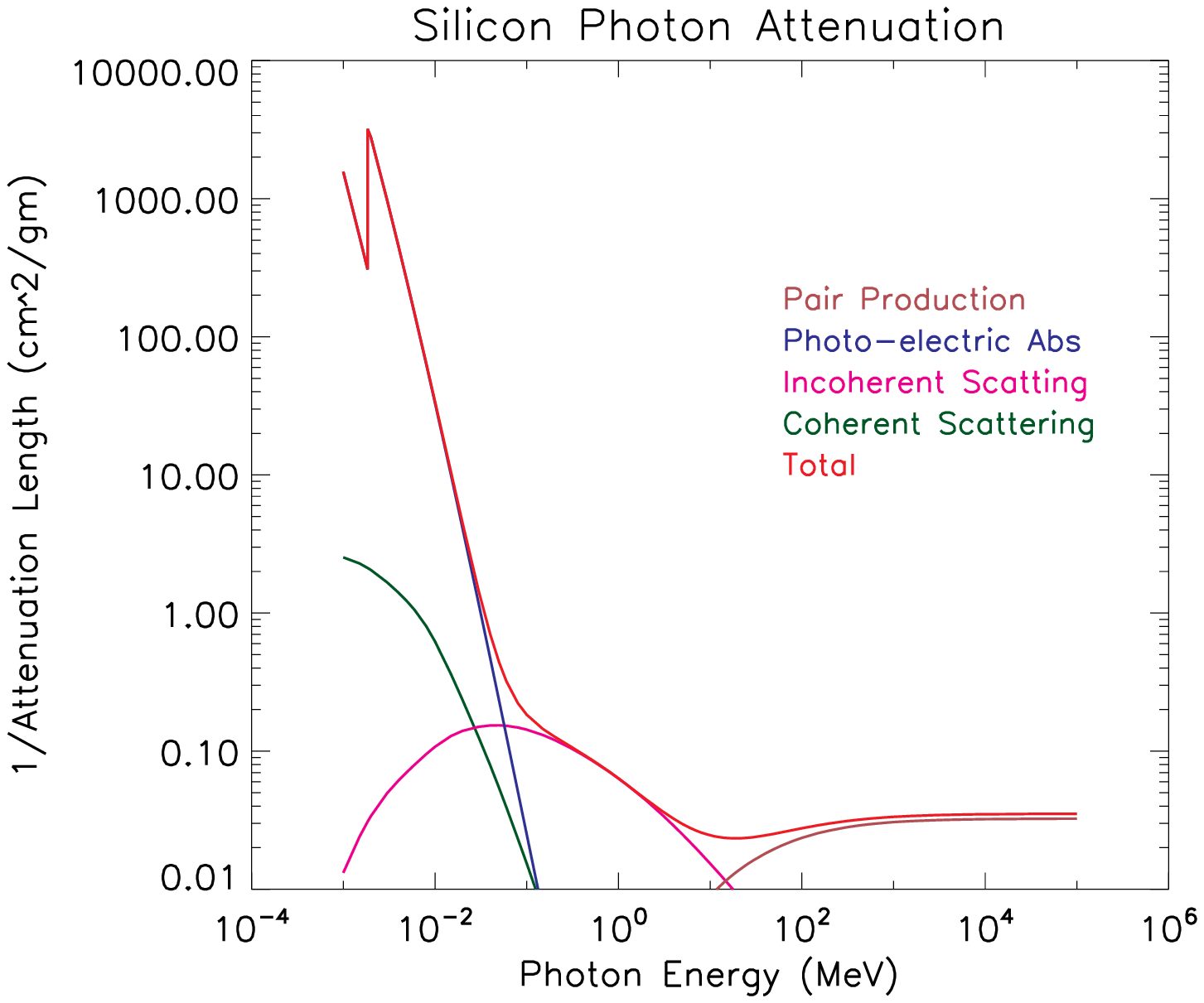}{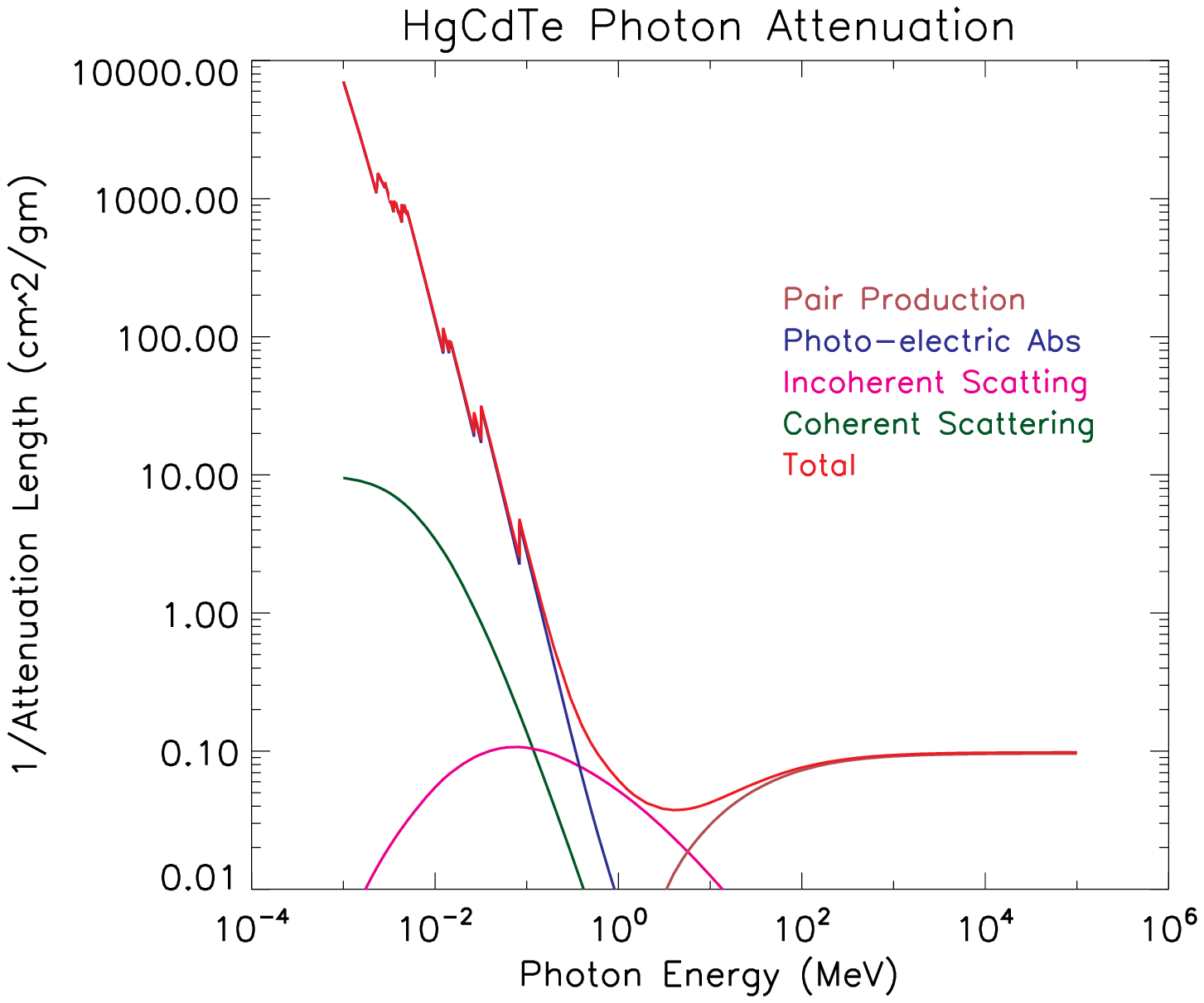}
\caption{The inverse attenuation lengths for photons in Silicon (left) and HgCdTe (right) are shown as a function of photon energy. All of the interaction processes other than coherent scattering generate energetic electrons within the detector material, in turn generating electron-hole pairs by means of ionization. The majority of the bremsstrahlung photons produced within the satellite that propagate any great distance will have energies of tens to hundreds of keV, where the photo-electric cross section is large. Thus, the probability of interaction in detector material is not insignificant.}\label{fig-pe_abs}
\end{figure}

Interaction probabilities for bremsstrahlung photons in HgCdTe and Si detector material can be calculated from the inverse attenuation lengths, plotted  in Figure \ref{fig-pe_abs}. The cross-sections for these plots were taken from the XCOM database \citep{xcom10}, with a representative stoichiometry for HgCdTe material of Hg$_{0.25}$Cd$_{0.25}$Te$_{0.5}$. At typical bremsstrahlung energies of 50-100 KeV and typical detector thicknesses of 7 and 15 microns for HgCdTe and Si, respectively, the interaction probabilities are $\approx$2\% and 0.1\%. These probabilities are small, but the number of bremsstrahlung photons present may be large, so the net contribution to detector backgrounds can be significant. These probabilities were also calculated only for the active layer of detector material; the secondary electrons produced may penetrate farther, so conversions in other nearby materials should be included when modeling the background rates. 

The photo-electric absorption cross-section increases strongly as a function of atomic number (Z), so high-Z materials provide the most effective photon shielding per unit mass; lead and tantalum are common choices. 
Mass constraints will likely preclude complete shielding against bremsstrahlung backgrounds; instead, shielding will be tailored so that the photon interaction rate in the detector is some acceptable fraction of the Galactic cosmic ray rate; \eg\  10\%. In this case, photon interactions occur throughout the high-Z shielding, each of which produces an energetic electron. Events occurring near the inner edge of the material may generate electrons with sufficient energy to pass through the remainder of the material and then pass unimpeded to the detector. Thus a practical shielding concept incorporates 3 different layers: an outer low-Z material to stop the majority of the electrons in the external environment (a low-Z material is preferred to minimize bremsstrahlung production), a high-Z layer to attenuate the bremsstrahlung flux produced in the observatory and outer shielding materials, and an inner low-Z layer to stop secondary photo-electrons produced at the inner edge of the middle layer. 

Fluorescence of X-rays in the high-Z shielding material is an important consideration in the overall design. When a high-energy photon ejects a photo-electron from an atom in the shielding material, the atom is left in an excited state, which decays by emission of a photon, a process known as  fluorescence. This is typically a \kalpha\ photon or ($\approx$20\% of the time) a \kbeta\ photon, which for lead have energies of 74.9\,keV and 84.9\,keV, respectively. The mean interaction length of a \kalpha\ photon in lead is 0.34\,mm, so a reasonable fraction of bremsstrahlung  absorption events in the inner 0.34\,mm of a lead shield will result in escape of a \kalpha\ photon into into the interior of the instrument. As can be seen in Figure \ref{fig-pe_abs}, such photons will have a greater detection probability than the higher-energy photon that was stopped in the lead. Thus it is possible for a thin high-Z shield to {\it increase} the photon-induced background in the detectors rather than decrease it. This shielding layer must be chosen to be thick enough that the reduction in incident bremsstrahlung flux outweighs the generation of fluorescent photons, after accounting for the difference in detection probabilities in the detector material.

Because the components of the instrument and spacecraft provide both some shielding and concomitant production of bremsstrahlung, and because the processes involved in both shielding and secondary particle production are strong functions of both material and  particle energy,  the shielding design will have to be optimized for each mission. An example of this optimization process will be given in Section \ref{sec-wf} below.

\subsubsection{\cerenkov\ Radiation}
The relativistic velocities of the electrons leads to yet another potential instrumental background: \cerenkov\ radiation. If a charged particle is passing through a refractive material with refractive index n($\lambda$) at a velocity exceeding c/n($\lambda$), it will emit photons, an effect called \cerenkov\ radiation. For a typical optical material with n($\lambda$) = 1.5, the critical velocity is $v_{c}/c = \beta_{c}$ = 0.667, which corresponds to an electron kinetic energy of 0.175\,MeV. The photons are emitted in a cone with half-angle $\theta_{c}$ relative to the particle's direction of motion, 
where $cos \theta_{c} = 1 / n(\lambda)\beta_{c}$. 

The energy lost through this radiation is small in comparison with the losses to ionization, but these radiated photons may be a significant background. The number of such photons emitted per unit path length and unit wavelength interval is given by:
\begin{equation}
\frac{d^{2}N}{dx d\lambda} = \frac{2 \pi \alpha z^{2}}{\lambda^{2}} 
\left ( 1 - \frac{1}{\beta^{2} n(\lambda)^{2}} \right )
\end{equation}
where $\alpha$ is the fine structure constant. 

For purposes of computing the emergent intensity per unit solid angle, the equation above should be evaluated at vacuum wavelengths, and then reduced by a factor of $1/n^2$ to account for the fact that the \'{e}tendue of an optical system at a given position in the optical path scales as $n^2$. 

\begin{figure}[htb!]
\epsscale{1.0}
\plotone{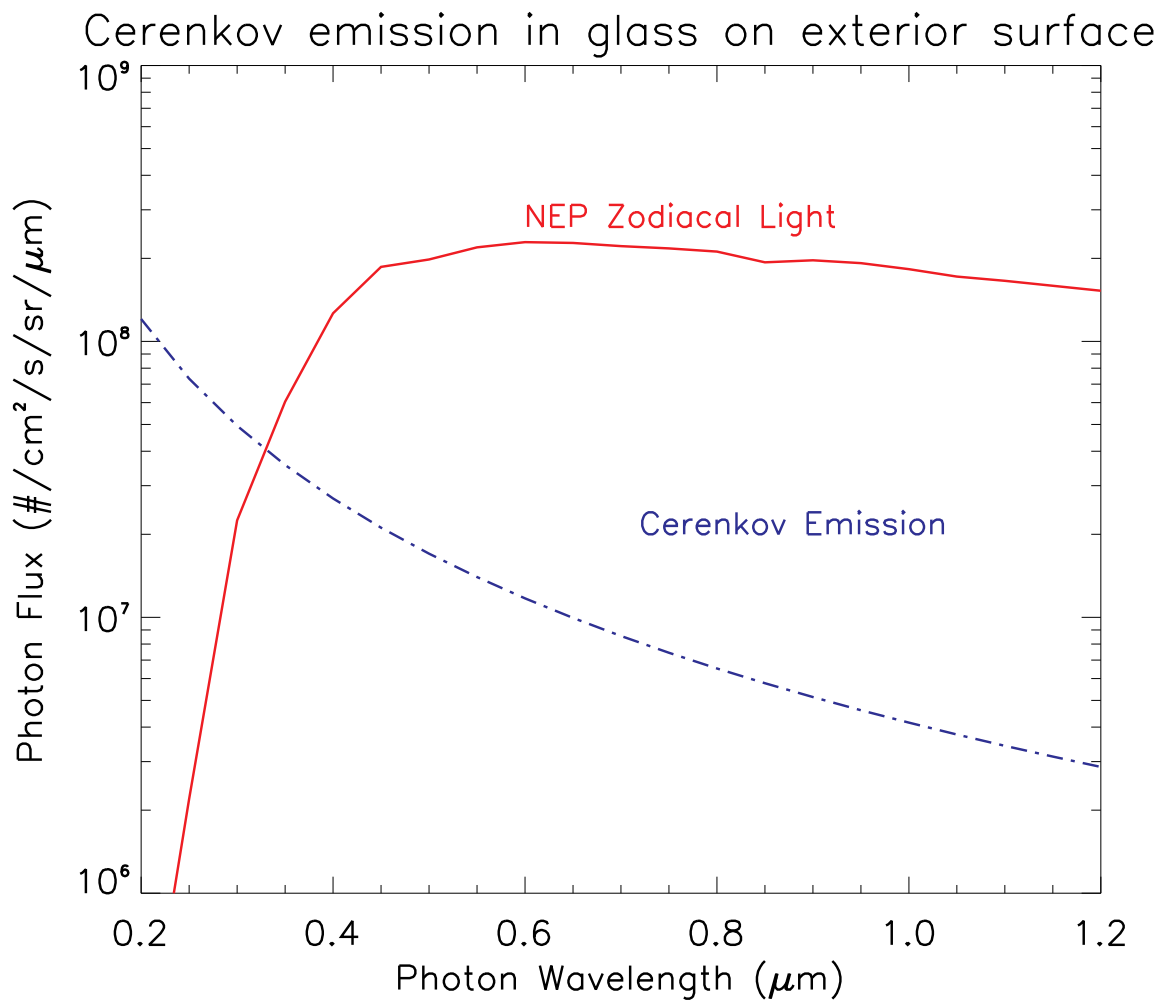}
\caption{The intensity of \cerenkov\ radiation in an optical element exposed to the unshielded electron environment in Geo-synchronous orbit is shown. Zodiacal light emission at the North Ecliptic Pole is shown for comparison, as it is the dominant astronomical background for many applications. \cerenkov\ emission dominates only at UV wavelengths. This comparison is meaningful for the case of refractive instruments where all optical elements but the first are well-shielded. If downstream optics are exposed to even a modest fraction of the external electron flux, especially if downstream from filters, then \cerenkov\ radiation may be a significant source of background.  }\label{fig-cer}
\end{figure}

The potential impact of \cerenkov\ radiation will be illustrated for two instrument geometries. The first case is a refractive camera with the first optical element at the exterior surface of the spacecraft. It is assumed that this optical element is exposed to the full electron flux shown as the solid line in Figure \ref{fig-AlBox}. The electrons in each energy bin were propagated in short steps through the glass, taken to be sapphire for the sake of concreteness, with the \cerenkov\ emission spectrum calculated at each step according to equation 1 above, until their energy fell below the threshold for emission. The cumulative spectrum is shown in Figure \ref{fig-cer}. The \cerenkov\ emission from any single electron is directional, as described above, but as the incident electron flux is isotropic, the cumulative \cerenkov\ emission will be as well.  
As this flux will appear as a diffuse background glow, zodiacal light emission is also shown in the Figure for comparison. Zodiacal light dominates for wavelengths longward of 0.4$\mu$m, but at shorter wavelengths the \cerenkov\ emission is brighter. One obvious implication of this figure is that filters employed in the optical system should be placed in locations well-shielded from the electron environment, as the detector will see the full \cerenkov\ spectrum emitted from materials downstream of the filter.

A second case of interest is a window placed in close proximity to a detector surface. This might be encountered, for example, in a reflective telescope with a CCD in a dewar at the focal plane. There would be an open path from outside of the observatory through to the dewar window. The electron environment for such an instrument may be similar to that of the 'open box' geometry described at the beginning of this section. The detector window would serve both to close out the thermal environment of the dewar and to prevent residual electrons from directly impacting the detector. The calculation in this case is similar to that in the preceding paragraph, except that the incident electron distribution was taken as the ``5mm wall'' electron spectrum shown as the red dashed line in Figure \ref{fig-AlBox}. For the sake of concreteness, the detector was taken to be the ``blue-optimized'' version of the e2V model 203 CCD. Assuming the window is in reasonably close proximity to the detector, the resulting detected \cerenkov\ emission count rate is 1.04$\times 10^{5} \: {\rm cm^{-2}s^{-1}}$. This CCD has 12$\mu$m pixels, which corresponds to 6.9${\rm \times 10^{5} \: pixels \,cm^{-2}}$, so the net background is 0.15 \cps. This far exceeds the dark current in the CCD, and may or may not be comparable to the sky background depending on the plate scale and other aspects of the instrument design. 

The two examples above demonstrate that \cerenkov\ radiation may be a significant instrumental background resulting from the energetic electron component of the Geo-synchronous orbit environment.

\section{The \wfirst\ Wide-Field Instrument} \label{sec-wf}

This section is organized into three subsections. The first provides a brief description of the \wfirst\ mission, the basic characteristics of the observing program, and the resulting requirement on the radiation background event rate. The second subsection provides a preliminary assessment of the radiation environment within the wide-field instrument in the absence of any dedicated shielding. This establishes the range of parameter space to be explored for the detailed shielding design, and in particular to determine a location for an optical window to block electrons from propagating along the light path. 
As will be seen: the required mass of shielding material can be substantial, so a careful analysis of the allowable radiation backgrounds is a critical component of optimizing the mission design. The final subsection describes the detailed calculation and results.

\subsection{The \wfirst\ observing program and radiation background requirements}

The Wide-Field Instrument (WFI) in the \wfirst\ observatory will be used to survey large areas of the sky, observing faint objects such as high-redshift galaxies. The \wfirst\ mission and instruments are described by \citet{sdt15}, and additional information on the WFI can be found in \citet{dac13}. Unlike the observing programs of the \sdo\ mission,  the majority of \wfirst\ WFI observations will be sky-background-limited and long exposures are required to achieve the desired signal to noise.  Therefore, backgrounds arising from the charged particle environment can seriously degrade the data. A final decision has not been reached for the orbit of the \wfirst\ observatory: both Sun-Earth L2 and Geo-synchronous orbits are presently under consideration. The energetic proton environment is similar for those two orbits, but the electron environments are very different. This section describes an initial assessment of shielding to limit the impact of the electron environment in Geo-synchronous orbit on the data.  

The \wfirst\ Project has not yet formally established criteria for the confidence levels to assign to ``acceptable'' performance. Therefore, we have adopted the 50\% confidence level Solar Maximum proton energy distribution (as shown in Figure \ref{fig-swp}), and the AE9 electron annual average energy distribution for purposes of the calculations presented here, with the understanding that shielding design may have to be augmented once the Project sets formal performance requirements. 

The \wfirst\ geo-synchronous orbit inclination is planned to be 28$\degr$, but all the calculations presented here were performed with the orbit environment in the equatorial plane, which corresponds to an orbit inclination of 0$\degr$. This choice was made because a significant fraction of each orbit would be spent near the equatorial plane, and it would not be practical to discard data obtained during those times. Thus the shielding would have to be adequate for that environment.

Assessment of radiation-induced effects on the data quality were assessed given the presence of other sources of backgrounds and data loss, as reducing radiation effects to levels much below those of other backgrounds brings decreasing benefits at possibly large increases in cost. The other backgrounds include uniformly-distributed noise sources such zodiacal light, detector dark current, telescope thermal emission, and detector readnoise, and localized sources of data loss such as bad pixels and gaps between detectors. Radiation-induced backgrounds can likewise be categorized as uniform (\cerenkov\ emission) or localized (energetic charged particle interactions in detectors). These will each be evaluated in detail below.

The \wfirst\ mission will execute a number of different observing programs, each with different requirements on limiting sensitivity, region of sky, and observing cadence, and with different mixtures of background sources. For the sake of concreteness, the imaging portion of the High-Latitude Survey \citep{sdt15} will be used here; calculations based on the other observing programs lead to similar conclusions. This program surveys over 2200 square degrees of sky at high ecliptic and Galactic latitudes in four filters spanning 0.92 -- 2.0 $\mu$m, to depths of 25.8 -- 26.6 magnitudes (AB). One of the objectives is to determine the distribution of dark matter by means of weak gravitational lensing of galaxies, which requires accurate measurements of the shapes of galaxies.  The theory underlying the dither strategy chosen for this survey that provides sampling of the PSF suitable for the weak lensing measurements is described by \citep{rhr11}. This analysis includes a means for quantitative assessment of the quality of the reconstructed image in the presence of bad pixels and/or pixels lost to cosmic rays. A comparison of the galaxy size distribution on the \wfirst\ pixel scale led to the adoption of the criterion that a cosmic-ray track passing within 3 pixels of the center of a galaxy would lead to exclusion of the exposure when combining the data for that galaxy. If a larger fraction of the pixels are affected by cosmic rays, a number of options can be considered: increase the shielding (if practical), increase the number of exposures on each region of sky so that there are enough unaffected exposures at fixed S/N for each galaxy, and increase the duration of exposures rather than their number,  so that the increased number of fainter galaxies  makes up for those lost to cosmic rays. The latter two options decrease the area of the sky that can be surveyed for a given allocation of observing time. The considerations involved in such a trade are complex and unique to each science program; 
an assessment of the baseline dither strategy for the high latitude imaging survey concluded that acceptable results would be obtained if fewer than 10\% of the galaxies were affected by cosmic rays (or instrumental effects such as bad detector pixels) in any single exposure.  
A plausible allocation of 1\% for the fraction of inoperable detector pixels leaves 9\% pixel loss per exposure as a design guideline for radiation shielding. 

Zodiacal light, the scattering of Sunlight by interplanetary dust (and thermal emission by the dust at wavelengths longward of 3.5$\mu$m), is an irreducible source of background, and is often the dominant source of background for space-based astronomical observatories located in the ecliptic plane at a distance of approximately 1\,AU from the Sun. As a point of reference, typical \wfirst\ observations planned for surveys of the extragalactic sky will have zodiacal light backgrounds of 0.25 -- 0.5 \cps, depending on the choice of bandpass filter and scheduling constraints. Dark current in the \wfirst\ detectors will be far lower than this, as will telescope thermal emission for all but the longest-wavelength filter. 
For the planned exposure time of 174 seconds, detector readnoise is roughly comparable to the Poisson noise on the backgrounds noted above, and a representative value for the total estimated noise is 14 electrons rms per pixel.

\cerenkov\ radiation produces a diffuse distribution of background photons within the instrument at wavelengths within the instrumental bandpass, hence these contribute to image noise in the same manner as zodiacal light.  \cerenkov\ photons produce single photoelectrons in the detector and cannot be distinguished from the astronomical photons of interest. Limits on \cerenkov\ radiation backgrounds are set by requiring this background to be ``small'' with respect to the zodiacal light, so as to not significantly increase the exposure time needed to reach a given signal-to-noise. A nominal requirement on \cerenkov\ radiation background is that it be less than 50\% of the zodiacal light background: given the present estimated contribution of readnoise, this level of \cerenkov\ emission results in a 10\% increase in the total noise for a pixel.  

Unlike \cerenkov\ photons, photoelectric conversion of high-energy bremsstrahlung photons and interactions of energetic charged particles in detector pixels deposit thousands of electrons in one or more pixels, typically overwhelming the integrated signal from faint galaxies. Hence, pixels affected during any one exposure are effectively equivalent to dead pixels from the standpoint of survey design. As can be seen from Figures \ref{fig-swp} and \ref{fig-AlBox}, the integral energy distribution is flat below 10\,MeV for protons and below 0.5\,MeV for electrons, for any appreciable shielding thickness. Thus the vast majority of the charged particles that penetrate shielding material have kinetic energy high enough for the range to far exceed the thickness of surface coatings, electrode structures, and depletion depth of the detector, hence all must be included in the 9\% pixel loss budget given above. The same applies for the photo-electric conversion events, as the majority of the bremsstrahlung and the X-ray fluorescent photons have sufficient energy to deposit a large signal. The shape of the energy distribution of energetic particles that penetrates shielding varies slowly with shielding thickness, so the primary effect of changing the shielding design is simply to change the rate of these background events in the detector.    

Because the affected pixels are highly localized and contain a signal exceeding the background level by well over 10-$\sigma$, they are readily identified by comparing  multiple images of the same field and may be discarded. Thus the impact on the data is loss of an exposure for the affected star or galaxy, rather than corrupted data. An exception is the case of an interaction that coincides with the position of a moderately bright star; in this case it may not be possible to identify the affected exposure and the impact is increased uncertainty in the characteristics of the stellar image (signal level, position, shape of the derived point-spread function, etc). 

Cosmic rays often interact with multiple pixels as they pass through a detector. The solid state detectors planned for use in the \wfirst\ Wide-Field Imager have depths of their active regions only slightly smaller than the transverse dimensions of the pixels (\eg\ $\approx$ 7$\mu$m depth and 10$\mu$m transverse). This means that cosmic rays have a significant probability of interacting with several adjacent pixels as they pass through a detector. The likely distribution of proton track lengths was estimated based on experience with the detectors on the Hubble Space Telescope (HST). The CCDs in the Wide-Field Channel in the Advanced Camera for Surveys (ACS) have nearly cubical pixels, with transverse dimensions of 15\,$\mu$m and a depth of 14--16\,$\mu$m \citep{mc98} and exhibit track lengths that typically range from 2 to 12 pixels (90\% of tracks have lengths of 12 pixels or fewer), a median track length of 5 pixels, and a tail extending out to 20 pixels and beyond \citep{AR02}. % no matching reference for WFC3, so I did the analysis myself using CANDELS data I had on hand.
At the other extreme is the HgCdTe detector in the HST WFC3 IR channel, which has pixels with transverse dimensions of 18\,$\mu$m and a depletion depth of only 6--8\,$\mu$m. Cosmic ray tracks in this detector are almost all only 1-2 pixels in length, with only 5\% percent being longer than 4 pixels. As the aspect ratio of the \wfirst\ detector pixels is midway between these limiting cases, we have adopted a track length of 4 pixels for the purposes of this paper. Combining the half-length of such a track with the 3-pixel radius threshold noted above for the effective galaxy size, gives a net footprint per cosmic ray event of 41 pixels. There are 10$^{6}$ pixels cm$^{-2}$ in the \wfirst\ NIR detectors, so combining this with the 174\,s exposure time and the 9\% allocation to cosmic-ray losses gives a maximum allowed cosmic-ray rate within the shielding of 12.6 \evts.

The near-infrared detectors planned for the \wfirst\ wide-field instrument can be read out non-destructively, unlike CCDs. A common use of this capability is to read out the detectors periodically throughout the exposure, and to determine the signal by fitting a linear slope to the series of readouts. One benefit of this approach is a significant increase in dynamic range for bright objects that would saturate the full-well depth, as readouts obtained prior to saturation are available for measuring the signal.  In principle, the series of readouts also enables detection and removal of cosmic rays, as they will appear as a one-time increase in the apparent count rate in a pixel. This one readout can be discarded, and the true signal determined from fitting the slope to the readouts before and after the one affected by the cosmic ray. This type of data processing is now routine for the HST/WFC3-IR detector, for example. It is not a panacea, however. For the relatively short exposures to be employed by \wfirst, the uncertainty on the derived count rate is significantly larger than for an unaffected exposure.  An additional complication is that these detectors exhibit persistence (also called a latent image), so the charge deposited by the cosmic ray will affect the signal measured in readouts following the cosmic ray hit. This effect is small and can be calibrated, but complete removal with the required accuracy may be difficult. Finally, for bright objects such as stars, the charge deposited in a pixel by a cosmic ray may not be large relative to the stellar signal, in which case robust detection of the cosmic ray may not be possible. For the initial \wfirst\ shielding assessment, it was deemed preferable to keep the majority of the radiation background from reaching the detectors at all, rather than rely on post-processing of the data. Hence for purposes of the analysis presented here, the shielding design was required to meet the event rate given above without any allowance for what might be recoverable from non-destructive readouts of the detectors.

\subsection{Radiation environment within instrument in absence of dedicated shielding}

\begin{figure}[htb!]
\epsscale{1.1}
\plottwo{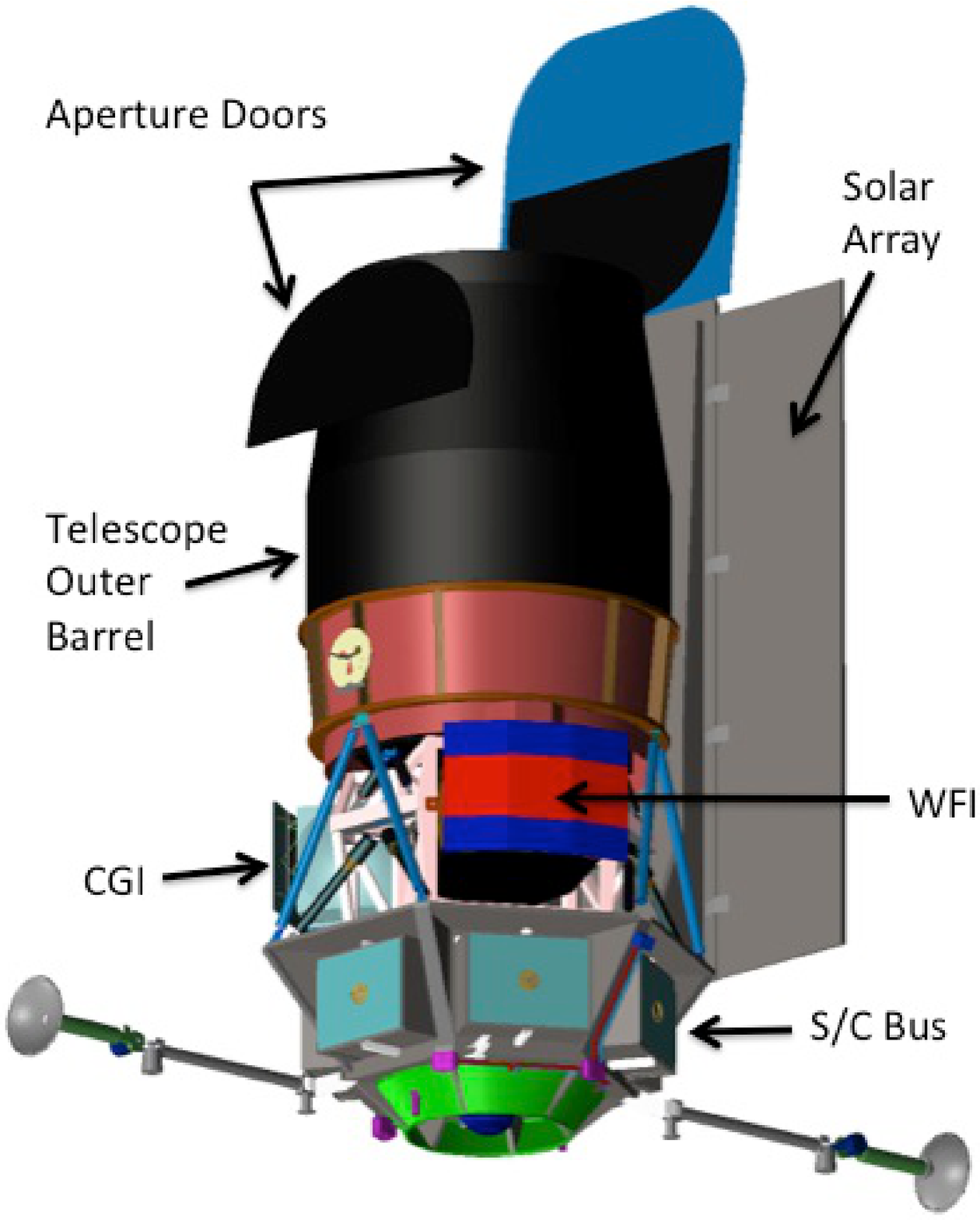}{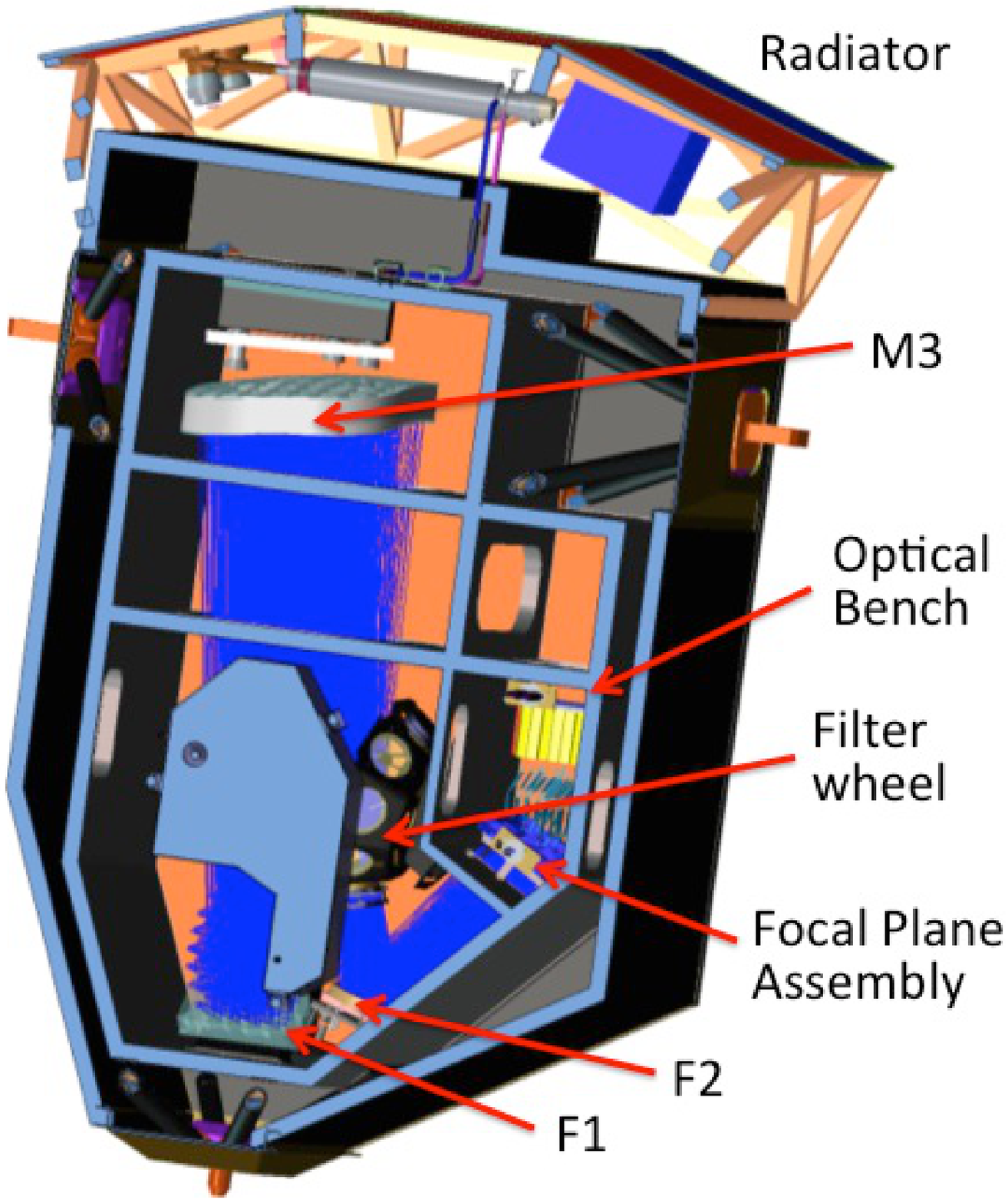}
\caption{CAD models of the \wfirst\ observatory (left panel) and the Wide-Field Instrument (WFI, right panel) are shown to illustrate the geometry of the particle shielding provided by the observatory and instrument structure. As can be seen in the left panel, the solid angle above and below the WFI is fairly-well blocked by high-mass components, while the areas to the sides are mostly open, with just a small fraction of the area blocked by structural elements.  
The light path in the WFI is shown in dark blue. The 3-layer electron-photon shielding (not shown) encloses the light-path from the filter wheel to the fold mirror F2, to the focal plane assembly (FPA), and then encloses the FPA and readout electronics.
}\label{fig-obs_elements}
\end{figure}

As noted above, the majority of the Galactic cosmic-ray protons have energies that are too high for shielding to be practical; the same also applies to the high-energy tail of the Solar particle event protons, particularly at Solar Maximum.  However, as shown in Figure \ref{fig-swp}, the majority of the Solar proton distribution is at energies where shielding can have a significant impact. A preliminary assessment of the effects of energetic protons on the \wfirst\ High-Latitude Survey observing program was made as follows.
The basic features of the layout of the \wfirst\ observatory and Wide-Field Instrument are illustrated in Figure \ref{fig-obs_elements}.
Using the WFI CAD model and simplified observatory model described below, we estimated that roughly half of the solid angle was shielded by the equivalent of 1.5\,cm of aluminum, and the remaining half of the solid angle evenly split by the equivalent of 3.5\,cm and 7.0\,cm of aluminum. With this level of shielding, the protons of Solar origin reaching the detector during Solar Maximum outnumber the Galactic cosmic ray protons reaching the detector at Solar Minimum (these two sources of backgrounds peak at opposite phases of the Solar cycle), so the Solar Maximum environment was used to evaluate the background event rate.  Propagating the 50\% CL Solar Maximum and GCR Solar Max proton spectra from Figure \ref{fig-swp} through this material, 
gives 5.5 \evts.  
This is well below the total budget of 12.6 \evts\ described above, leaving 7.1 \evts\ as an allocation for losses due to the electron environment, assuming the same footprint per event as for energetic protons. Photon interactions in detector pixels generally produce secondary charged particles with a range smaller than a pixel, and most of the charge is deposited in one, or at most two, pixels. 

For the initial assessment of the electron environment within the instrument, we simply assumed that the telescope and spacecraft were opaque over half of the solid angle, and that the remainder of the solid angle was dominated by the composite panels enclosing the WFI. The first question addressed was whether or not it would be acceptable to have a window immediately in front of the detector for purposes of keeping the residual electron flux from reaching the detectors. The WFI instrument includes filters spanning the range 0.76--2.0\,$\mu$m, so we assumed that a filter with this wide bandpass would be applied to the back surface of this detector window to limit the \cerenkov\ emission that would reach the detectors. With the level of shielding described above, the electron flux incident on the window was roughly 5 times higher than that shown for the 5\,mm aluminum wall case in Figure \ref{fig-AlBox}. After scaling for the fact that the \wfirst\ detectors have 10$\mu$m pixels, the net \cerenkov\ radiation background was 0.68 \cps. This was considered unacceptable, as this would represent an increase over the zodiacal light background by factors of two to three, in comparison with our nominal requirement for an increase of no more than 50\%. The conclusion was that filters located in the light path at the intermediate pupil upstream of the the detector would serve to block electrons from propagating along the light path, and that shielding would be added to enclose the light path between the detectors and the filters. The filters are sufficiently far from the the detectors that the \cerenkov\ radiation emitted within the filters would produce very little background.

After determining that the filters would be the best location for blocking the passage of electrons along the light path, several combinations of multi-layer shielding were considered for the enclosure. In all cases the outer and inner layers were graphite epoxy with the center layer being lead. Thicknesses investigated ranged from 13-19\,mm for the outer layer, 1-2\,mm for the lead layer, and 3-4\,mm for the inner layer. 

\subsection{Detailed shielding design}

In the next iteration of the design, propagation of the radiation environment through the instrument structure was modeled in detail with the NOVICE code \citep{Jor76, Des10}. The full CAD model of the instrument was incorporated, including the properties of the materials. To keep the computation time manageable, simplified geometric models of the telescope and spacecraft bus components were employed, though still with realistic distributions of mass and materials. 

The present version of the NOVICE code includes energy deposition from absorption of X-ray fluorescence photons, but does not provide information on the propagation of those photons in its output. In order to estimate this contribution to the total detector background, we modeled the generation and propagation of these photons in the following manner.  
The NOVICE model of the bremsstrahlung photon spectrum gives the photon fluxes as a function of energy at the top and bottom surfaces of the lead layer, and we assumed that the flux distribution varies linearly through the thickness of the lead (a reasonable approximation for the thicknesses of interest here): 
\begin{equation}
N(E,z) = N_{bot}(E) + (N_{top}(E) - N_{bot}(E)) \times z / T,
\end{equation}
where $T$ is the thickness of the lead.
The volumetric emission density $J_{i}(z)$ in ${\rm photons\:cm^{-3}\:s^{-1}}$ for each of the \kalpha\  and \kbeta\ lines is given by:
\begin{eqnarray}
J_{i}(z) & = & \int_{K_{edge}}^{\infty} N(E,z)\ \rho\ \tau_{PE}(E)\ f_{PE}(K)\ \omega_{K}\ f_{i}\ dE \\
%J_{i}(z) & = & J_{i0} + (J_{iT} - J_{i0})  \ \frac{z}{T}  \nonumber \\
%J_{i}(z) & = & J_{i0} \left (1 + \frac{\Delta J_{i}}{T} z \right ), 
J_{i}(z) & = & J_{i0} \left (1 + \Delta J_{i} \frac{z}{T} \right ), 
\ \ \ \ \ \Delta J_{i} = \frac{J_{iT} - J_{i0}}{J_{i0}}.  \nonumber 
\end{eqnarray}
The factors in the integrand are the density of the material ($\rho$), the inverse attenuation length in cm$^{2}$/gm for photo-electric absorption as a function of energy ($\tau_{PE}(E)$), the fraction $f_{PE}(K)$ of the photo-electric absorption resulting in excitation of a K-shell electron, the fraction of K-shell de-excitation giving rise to a fluorescent photon ($\omega_{K}$), and the fraction of these photons going into a particular \kalpha\ or \kbeta\ line ($f_{i}$). $\tau_{PE}(E)$ was taken from the XCOM database \citep{xcom10}. The fluorescent photon energies and fractions $f_{i}$ were taken from the X-ray Data Booklet \citep{Thompson09}. $f_{PE}(K)$ was calculated to be 0.83 (for a hydrogenic atom well above threshold; at threshold this can be calculated directly from the XCOM cross-sections to be 0.79). Finally, $\omega_{K}$ was taken to be 0.9634 from \citet{Hub94}.

The volumetric emission density can then be integrated over angle and path length to obtain the emergent flux from the inner surface of the lead layer.  
The flux in a given fluorescence line $\Phi_{i}$ emerging from a unit area dA on the surface is obtained by computing the emission from a point P, applying attenuation along the path length r, and integrating over all angles $\theta$, $\phi$, and path lengths $r$ from 0 to $r_{max} = T/cos\:\theta$:
\begin{displaymath}
\Phi_{i} = \int_{0}^{2\pi}d\phi \int_{0}^{\pi/2} d\theta \int_{0}^{r_{max}} dr\, sin\theta\, cos\theta\,
e^{-r/L_{i}}\, \frac{J_{i0}}{4\,\pi} \left ( 1 + \frac{\Delta\:J_{i}}{T}\:r\,cos\theta\ \right ),  
\end{displaymath}
where  $L_{i}$ is the attenuation length in lead for emission line $i$. This expression reduces to:
\begin{equation}
\Phi_{i} = \frac{L_{i} J_{i0}}{2} \left ( \frac{1}{2} - ( 1 + \Delta J_{i}) E_{3}(\beta) +
\frac{L _{i}\Delta J_{i}}{T} ( \frac{1}{3} - E_{4}(\beta) ) \right ),
\end{equation}
where $\beta = T/L_{i}$, and $E_{n}(x)$ is the exponential integral:
\begin{displaymath}
E_{n}(x) = \int_{1}^{\infty} \frac{e^{-ux}}{u^{n}} du.
\end{displaymath}

\begin{figure}[tbh!]
\epsscale{0.9}
\plotone{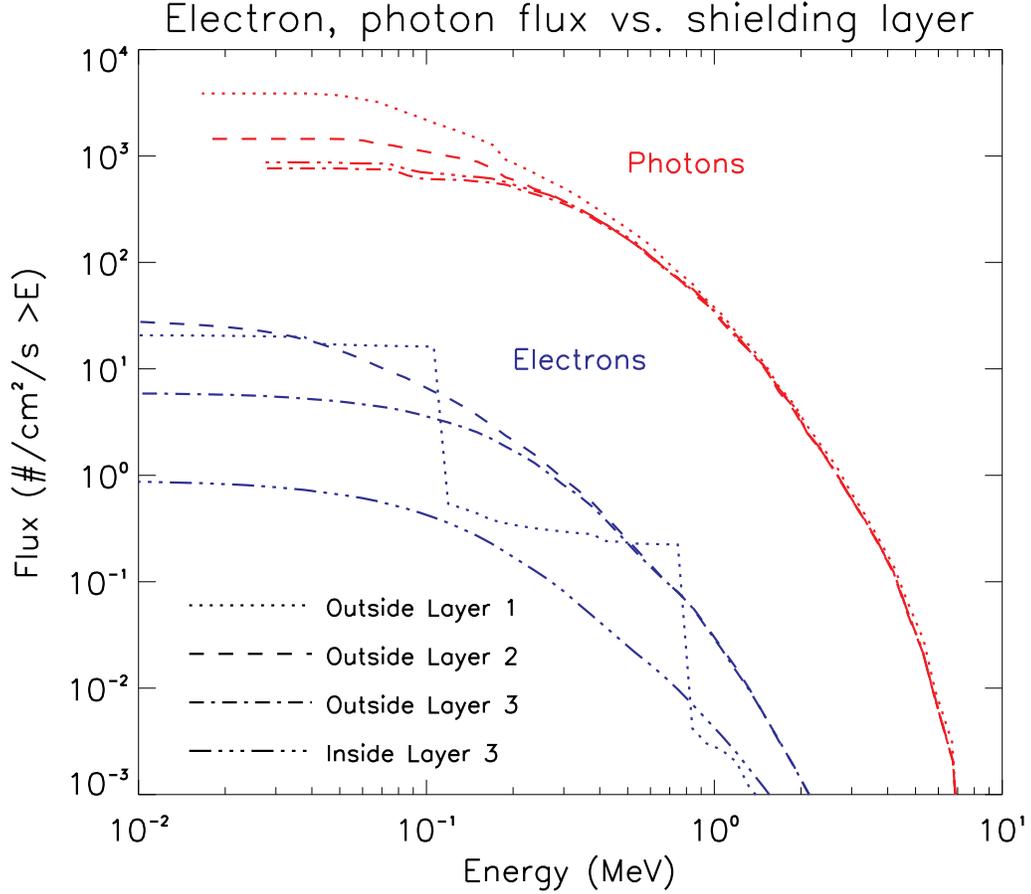}
\caption{The integral electron and photon flux distributions are plotted for various positions within the \wfirst\ Wide-Field Imager instrument. 
Electrons are plotted in blue, photons in red. The dotted lines give the flux inside the instrument after attenuation by observatory structure, but outside of the dedicated shielding enclosing the light path and detectors. The dedicated shielding consists of 3 layers: layers 1 \& 3  are 12\,mm and 3\,mm of graphite-epoxy, respectively, and layer 2 is 1\,mm of lead. It is interesting to note that the electron flux is higher between layers 1 and 3 than outside layer 1: this is caused by photo-electric conversion of bremsstrahlung produced in the observatory structure. The net electron flux within the shielding layers is only 1 \pden, which is below the Galactic cosmic-ray rate of 1.6--4 \pden.
The shielding reduces the photon flux by approximately a factor of 5. This photon flux produces an event rate in the WFI HgCdTe detectors of $\approx$ 4.9\cps. This is comparable to the energetic proton event rate, which is within requirements but just high enough for continued optimization of the shielding design to be worthwhile.}\label{fig-eshld}
\end{figure}

Several combinations of shielding wall thicknesses were evaluated. Results for the case of 12\,mm, 1\,mm, and 3\,mm thicknesses for the outer graphite-epoxy, lead, inner graphite-epoxy walls, respectively, are shown in Figure \ref{fig-eshld}.  
Fluxes are for particles crossing the side of the shielding enclosure within the \wfirst\ Wide-Field Imager instrument (WFI), at various depths. 
The dotted lines give the flux inside the instrument after attenuation by observatory structure, but outside of the dedicated shielding enclosing the light path and detectors. 
It is interesting to note that the total electron flux is actually higher between layers 1 and 3 than it is outside layer 1: this is caused by photo-electric conversion of bremsstrahlung produced in the observatory structure. The net electron flux within the shielding layers is only 1\pden, which is well below the energetic proton rate outside the shielding of 5.5\pden.

Figure \ref{fig-eshld} also shows the integral photon energy distribution for this shielding configuration. This includes both the bremsstrahlung spectrum computed by the NOVICE code and the \kalpha\ and \kbeta\ fluorescence photons computed from Equation 5. The shielding has little effect on the number of high-energy photons, but they are relatively few and have the lowest interaction probability in the detector material. The inner graphite-epoxy layer of shielding does not significantly attenuate the fluorescent X-ray photons produced in the lead layer. The photon flux shown is the total particle flux crossing the surface: it includes particles entering the surface as well as those exiting. This is the desired quantity when calculating the fluorescence yield, but must be kept in mind when calculating the flux crossing the detector surface to avoid double-counting. In particular: the quantity $\Phi$ calculated in Equation 5 is the flux per unit area emitted into the detector enclosure, but does not include the flux emitted from the other walls of the enclosure that is crossing into the surface. Thus $\Phi$ was doubled to have the same units as the bremsstrahlung spectrum when computing the net photon spectrum in Figure \ref{fig-eshld}.

The photon background event rate $R_{\gamma}$ in the detector material is given by:
%\begin{displaymath}
\begin{equation}
R_{\gamma}  =  \sum_{K_{i}} \int_{0}^{2\pi}d\phi \int_{-\pi/2}^{\pi/2} d\theta \,sin\,\theta \, cos\,\theta \,
\int [ 1 - e^{-\rho\, \tau(E) T / cos\,\theta} ]\, \Phi_{i}(E,\theta,\phi) \, dE,
\end{equation}
%\end{displaymath}
where $\rho$ is the density of HgCdTe (7.07 gm/cm$^{3}$), $T$ is the thickness of the detector material, $\tau (E)$ is the total photon absorption coefficient in the detector material excluding coherent scattering, expressed as the inverse attenuation length in cm$^{2}$/gm (see the right panel of Figure \ref{fig-pe_abs}), $\Phi_{i}(E,\theta,\phi)$ is the particle flux per unit area per steradian in fluorescent  line $i$, and the sum runs over all of the \kalpha\ and \kbeta\ lines. The background resulting from bremsstrahlung applies if $\Phi$ is set equal to the bremsstrahlung spectrum. This expression simplifies considerably if we take $\Phi$ to be isotropic ( $\Phi_{i}(E,\theta,\phi) = \Phi_{i}(E)/ 2\pi$, with $\Phi_{i}(E)$ calculated as in Equation 5 above):
%\begin{displaymath}
\begin{equation}
R_{\gamma}  =   \sum_{K_{i}} \int [ \frac{1}{2} - E_{3}(\rho\,  T\, \tau(E))] \, \Phi_{i}(E) \, dE.
\end{equation}
%\end{displaymath}

Evaluating $R_{\gamma}$ for the flux outside layer 1 (what would be present in the absence of dedicated shielding) gives an interaction rate of 60 \evts, far too high to be acceptable. The event rate is only 4.9\evts\ for the photon flux inside layer 3, however, of which 1.6\evts\ result from fluorescence emission in the lead. 
When these photon-induced events are combined with the electron event rate of 1\pden, the total of 5.9\pden\ is within the design requirement of 7.1\pden noted at the beginning of this section. In practice, we do a bit better because $\Phi$ is not isotropic: the detector packaging, focal plane assembly, electronics, etc. mounted just behind the detectors reduces the photon flux reaching detector active layer. A rough estimate is that this would reduce the event rates above by $\approx$25\%. 
Finally,  the multi-layer electron-photon shielding reduces the proton event rate from 5.5\,\pden\ to 3.7\,\pden. 
The total modeled event rate then becomes 9.6\,\pden, which has $\approx$ 30\% margin against the requirement of 12.6\,\pden.

For reference: if the detector had been a silicon CCD with a depletion depth of 15 $\mu$m instead of HgCdTe, the photon event rates would have been roughly an order of magnitude lower. 

Substantial mass is required for shielding against the electron environment and concomitant bremsstrahlung photon background: the total mass of the shielding materials in the design presented here is $\approx$48\,kg, not including the increase in structural materials required to support it.  
The proton environment assumed for these calculations was the 50\% CL level for Solar Maximum, so it is not unlikely that additional shielding would have to be added to ensure that event rate requirements are met over the course of a mission lasting many years. 
This mass is significant even in the context of a large observatory, so careful consideration of the impact of radiation backgrounds on the data is warranted when setting requirements for shielding. It is possible that some mass savings can be achieved through a global optimization of component placement and material selection, but this analysis has not yet been attempted.
 
\section{Summary}
The radiation environment in Geo-synchronous orbit presents some unusual challenges to instrument design. In particular, the presence of high fluxes of relativistic electrons can create high backgrounds in common types of detectors that will degrade the quality of the data. In this paper we reviewed the basic characteristics of this radiation environment, and the physical principles governing the interaction of the radiation with structural, optical, and detector materials. Idealized enclosure cases were explored to illustrate how the magnitudes of the backgrounds can be affected by instrument design choices, and how these backgrounds can be substantially greater than the unavoidable astronomical backgrounds that one ordinarily considers when designing a mission. Finally, these concepts were applied in a realistic design exercise for the case of the \wfirst\ Wide-Field Instrument, and a plausible shielding design was presented that resulted in backgrounds induced by the Geo-synchronous orbit environment that were acceptable for meeting the mission scientific performance requirements.

\acknowledgments

The authors would like to thank the anonymous referee for numerous comments that improved the clarity of presentation throughout the paper.
The authors would like to thank Tom Jordan for numerous discussions on usage of the NOVICE code for our application, and 
one of us (JWK) wishes to thank Bernie Rauscher for a number of useful discussions on detector architecture.
This work made use of data on the Geo-synchronous orbit environment obtained by the EPEAD instrument on the GOES-13 satellite, and provided via the National Geophysical Data Center operated by NOAA. We made use of energy loss tables from the NIST Standard Reference Database 124 (\url{http://www.nist.gov/pml/data/star/index.cfm}) and photon cross sections from the NIST XCOM database (\url{http://www.nist.gov/pml/data/xcom/index.cfm}). This work was supported by the \wfirst\ Study Office at NASA/GSFC. C.H. is supported by the David and Lucile Packard Foundation, the Simons Foundation, and the U.S. Department of Energy.

\appendix

\end{document}